# Simulation of 10830 Å absorption with a 3D hydrodynamic model reveals the solar He abundance in upper atmosphere of WASP-107b

M. L. Khodachenko[1,2,3], I. F. Shaikhislamov[2], L. Fossati[1], H. Lammer[1], M.S. Rumenskikh[2,3], A. G. Berezutsky[2,3], I. B. Miroshnichenko[2], M.A. Efimof[2]

1) Space Research Institute, Austrian Academy of Sciences, Graz, Austria
2) Institute of Laser Physics SB RAS, Novosibirsk, Russia
3) Institute of Astronomy, Russian Academy of Sciences, Moscow, Russia

**ABSTRACT:** Transmission spectroscopy of WASP-107b revealed 7-8% absorption at the position of metastable HeI triplet at 10830 Å in Doppler velocity range of [-20; 10] km/s, which is stronger than that measured in other exoplanets. With a dedicated 3D self-consistent hydrodynamic multi-fluid model we calculated the expanding upper atmosphere of WASP-107b and reproduced within the observations accuracy the measured HeI absorption profiles, constraining the stellar XUV flux to (6-10) erg cm$^{-2}$ s$^{-1}$ at 1 a.u., and the upper atmosphere helium abundance He/H to 0.075-0.15. The radiation pressure acting on the metastable HeI atoms was shown to be an important factor affecting the shape of the absorption profiles. Its effect is counterbalanced by the processes of collisional depopulation of the HeI metastable state. Altogether, the observed HeI absorption in WASP-107b can be interpreted with the expected reasonable parameters of the stellar-planetary system and appropriate account of the electron and atom impact processes.

**Key words**: hydrodynamics – plasmas – planets and satellites: individual: exoplanets – planets and satellites: physical evolution – planets and satellites: atmosphere – planet–star interactions

## 1. Introduction

Following the study of *Seager & Sasselov (2000)*, *Oklopčić & Hirata (2018)* indicated that absorption of the near-infrared (NIR) stellar emission at the position of metastable helium 2$^3$S triplet at 10830 Å (hereafter HeI(2$^3$S)) enables an alternative way for probing the expanding exoplanetary atmospheres. Around the same time, *Spake et al. (2018)* observed with HST the primary transit of the warm Saturn-mass planet WASP-107b obtaining an excess 0.05% absorption in 98 Å range around the HeI(2$^3$S) triplet. Since this first detection, an excess absorption was measured for a number of close-orbit giant exoplanets, employing high resolution spectrographs that resolve the HeI(2$^3$S) triplet.

Two sets of ground-based transit observations of WASP-107b revealed 7-8% absorption, at the 20σ level, within [-20; 10] km/s velocity range around the HeI(2$^3$S) lines (*Allart et al. 2019*; *Kirk et al. 2020*). This absorption is significantly stronger than that of ~1% measured for the warm Neptunes HATP-11b (*Allart et al. 2018, Mansfield et al. 2018*) and GJ3470b (*Ninan et al. 2020, Pallé et al. 2020*), as well as for the hot Jupiters HD189733b (*Salz et al. 2018, Guilluy et al. 2020*) and HD209458b (*Alonso-Floriano et al. 2019*). It is also larger than ~3% absorption detected for another warm Saturn Wasp-69b (*Nortmann et al. 2018*). Based on the modelling done in *Salz et al. (2016)* and *Oklopčić & Hirata (2018)*, *Ninan et al. (2020)* and *Lampon et al. (2020)* derived a factor of 5-10 sub-solar He abundance values for GJ3470b and HD209458b, respectively (the solar value is He/H~0.1). Therefore, the deeper HeI(2$^3$S) transit detected for WASP-107b might be a sign of a solar, or even higher, He abundance. To explore this possibility, we run a set of self-consistent three-dimensional (3D) hydrodynamic (HD) multi-fluid simulations of the expanding and escaping upper atmosphere of WASP-107b, aimed at reproducing the measured HeI(2$^3$S) transit absorption. So far, the atmospheric outflow of WASP-107b has not been modeled with aeronomy codes, and the only available interpretation of the observations is that by *Allart et al. (2019)*, who employed a 3D Monte-Carlo approach. An important result of this study consists in estimating and taking into account of the effect of stellar NIR radiation pressure, acting on HeI(2$^3$S) atoms, which appeared ~75 times larger than stellar gravity. This led *Allart et al. (2019)* to artificially reduce the radiation pressure by two orders of magnitude to fit the observations.

## 2. The modelling approach

The 3D HD code, used in the present study, was previously employed for the modeling and interpretation of Lyα transmission spectroscopy observations of GJ436b (*Khodachenko et al. 2019*), HD209458b (*Shaikhislamov et al. 2020a*), π Men C (*Shaikhislamov et al. 2020b*), and GJ3470b (*Shaikhislamov et al. 2020c*). For GJ3470b, also the HeI(2$^3$S) absorption was simulated, revealing He/H=0.013, in agreement with *Ninan et al. (2020)*.

Note, that *Shaikhislamov et al.* (*2020c*) did not consider the stellar radiation pressure acting on the HeI($2^3$S) atoms, which is now taken into account in the reported modelling of WASP-107b.

The code solves continuity, momentum, and energy equations for all considered species, i.e., H, H$^+$, H$_2$, H$_2^+$, H$_3^+$, He, He$^+$, and He$^{2+}$. The metastable HeI($2^3$S) atoms are treated as a separate fluid with its own velocity and temperature, determined by those of the species from which they originate, namely He$^+$ or HeI($1^3$S), depending on whether recombination, or excitation from the ground state, respectively, generate HeI($2^3$S). Elastic collisions with other species also affect the macroscopic physical parameters of the HeI($2^3$S) fluid. All reactions, which populate and depopulate the HeI($2^3$S) component, are described in *Shaikhislamov et al.* (*2020b*), and are the same as those listed in *Oklopčić & Hirata* (*2018*).

Following *Allart et al. (2019)* we take as a proxy of stellar XUV emission at wavelengths λ<912 Å, the semi-synthetic spectrum of the analogous K6 star HD85512, provided by the MUSCLES survey (*France et al. 2016*). From this we derived an XUV flux $F_{XUV}$ ~3 erg s$^{-1}$ cm$^{-2}$ at the reference distance of 1 a.u. At longer wavelengths, we employed a synthetic stellar spectral energy distribution (SED), computed with the LLmodels stellar atmosphere code (*Shulyak et al. 2004*), considering the stellar parameters given in *Anderson et al. (2017)*. The photoionization of HeI($2^3$S) atoms by stellar near-ultraviolet (NUV) emission has a threshold at 2600 Å, and the employed stellar SED reveals in the range of 1000 Å < λ < 2600 Å at 1 a.u. the flux of $F_{NUV}$ ~15 erg s$^{-1}$ cm$^{-2}$. It yields the photoionization time of ~12 min at WASP-107b orbit. Another crucial value provided by the stellar SED is the NIR radiation flux around the HeI($2^3$S) triplet (10830 Å), which at 1 a.u. has a typical level of $F_{10830}$ ~11 erg s$^{-1}$ cm$^{-2}$ Å$^{-1}$. This NIR flux produces ~85 times larger than stellar gravity radiation pressure force acting on HeI($2^3$S) atoms.

Our model also calculates self-consistently the stellar wind (SW) plasma over the whole star-planet system, parametrized with the total stellar mass loss rate $M'_{SW}$, coronal temperature $T_{cor}$, and terminal SW speed $V_{SW,\infty}$ (see e.g. in *Khodachenko et al. 2019, Shaikhislamov et al. 2020a*). A set of simulation runs with a strong SW, assuming several times higher $M'_{SW}$ than that of the Sun (~2.5×10$^{12}$ g/s), revealed that HeI($2^3$S) absorption, which is produced relatively close to the planet, is not affected by interaction of the escaping atmosphere with the SW. Therefore, for simplicity's sake, we performed simulations under conditions of a weak SW, taking $M'_{SW}$~10$^{11}$ g/s, $T_{cor}$ ~ 10$^6$K and $V_{SW,\infty}$~400 km/s. The corresponding SW parameters at the planetary orbit are $V_{SW}$=200 km/s, $T_{SW}$=0.34 MK, $n_{SW}$=360 cm$^{-3}$.

| N | $F_{XUV}$ [erg cm$^{-2}$ s$^{-1}$]; $M'_p$ [×10$^{10}$g/s] | $F_{10830}$ [erg cm$^{-2}$ s$^{-1}$ Å$^{-1}$] | $A_{He,\%}$ max | $A_{He,\%}$ [-10; 10] | $A_{He,\%}$ [-40; -10] | $A_{He,\%}$ [10; 20] | He/H |
|---|---|---|---|---|---|---|---|
| 1 | 3 (7.6) | 11 | 5.7 | 4.3 | 1.8 | 0.2 | 0.1 |
| 2 | 6 (12.5) | 11 | 8.4 | 6.8 | 2.8 | 1.3 | 0.1 |
| 3 | 8 (15.3) | 11 | 9.4 | 7.9 | 3.5 | 2.3 | 0.1 |
| 4 | 10 (18) | 11 | 10.3 | 8.9 | 4.2 | 3.2 | 0.1 |
| 5 | 6 (12.5) | 5.0 | 7.5 | 6.4 | 1.9 | 4.9 | 0.1 |
| 6 | 6 (12.5) | 7.5 | 9.4 | 7.3 | 2.0 | 2.8 | 0.1 |
| 7 | 6 (12.5) | 15 | 7.4 | 5.6 | 3.4 | 0.6 | 0.1 |
| 8 | 3 (7.0) | 11 | 7.7 | 5.5 | 2.9 | 2.5 | 0.25 |
| 9 | 6 (11.8) | 11 | 9.8 | 8.0 | 3.5 | 1.5 | 0.15 |
| 10 | 10 (18) | 11 | 8.7 | 7.5 | 3.4 | 2.8 | 0.075 |
| *Allart et al. (2019)* | | | 10.0 | 7.2 | 3.5 | 3.6 | |
| *Kirk et al. (2020)* | | | 7.3 | 6.0 | 2.9 | 2.1 | |

**Table 1.** HeI($2^3$S) absorption values, simulated with different $F_{XUV}$, $F_{10830}$, and He/H in comparison to the observations (*Allart et al. 2019, Kirk et al. 2020*). 2$^{nd}$ column: Integrated XUV flux (10-912 Å) at 1 a.u. and corresponding planetary mass loss rate; 3$^{rd}$ column: NIR flux at the position of HeI($2^3$S) triplet at 1 a.u.; 4$^{th}$ column: absorption peak; 5$^{th}$, 6$^{th}$, and 7$^{th}$ columns: absorption averaged over Doppler velocity ranges [-10; 10] km/s (line center), [-40; -10] km/s (blue wing), and [10; 20] km/s (red wing), respectively; 8$^{th}$ column: helium abundance.

The model equations are solved on a spherical grid in the planet-centered reference frame with polar-axis Z, perpendicular to the orbital plane (see in *Shaikhislamov et al. 2020a*). For all model runs, we set at the base of planetary atmosphere a temperature 750 K and pressure 0.05 bar. While the currently unmeasurable $F_{XUV}$ is a free parameter of the model, the above specified stellar NIR and NUV fluxes are nearly constant and constrained by the stellar SED. However, to show the importance of radiation pressure for WASP-107b, we did several model runs with different $F_{10830}$. Table 1 presents the simulated HeI($2^3$S) absorption values obtained in 10 model runs with different $F_{XUV}$, $F_{10830}$, and He/H, and compares them with the observations.

## 3. Results

Figure 1 shows distribution of HeI($2^3$S) atoms in orbital plane (top) and the profiles of their temperature and velocity along Z axis (bottom), as well as density of major species, obtained in the model run N1, with the expected stellar flux values and solar He abundance.

Figure 2 shows the rates of considered reactions, responsible for the production and destruction of metastable HeI($2^3$S), as functions of distance from the planet. The main processes, balancing recombination pumping, are of collisional nature. They involve HeI($2^3$S) auto-ionization reactions with $H_2$ and H, which are important relatively close to the planet (<$2R_p$), and the electron impact transfer from the triplet to singlet state (reaction 3). Photoionization from the $2^3$S state is negligible everywhere, except of far region (>$10R_p$), where the density of species decreases significantly.

10830 Å triplet. The synthetic profiles were calculated for different values of $F_{XUV}$ (model runs N1, N2, N3, N4). To provide realistic comparison with observations, we average the simulated absorption over the planetary orbital phase interval ±0.0083 around mid-transit, which is between transit contact points II and III.

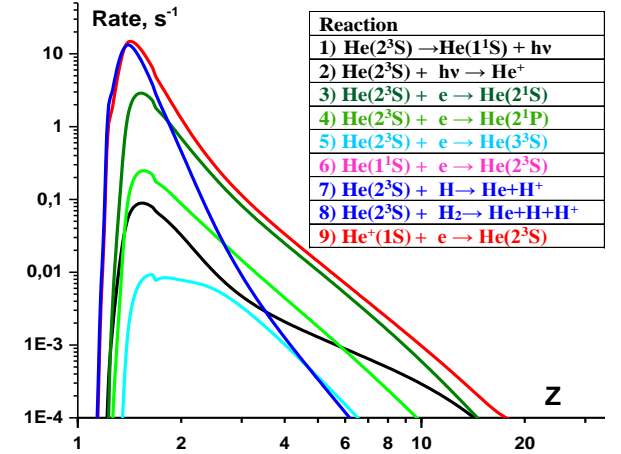

**Figure 2.** Rates of reactions responsible for the production and destruction of metastable HeI($2^3$S), versus distance along polar axis Z under conditions of the model run N1. Black line shows the sum of reactions 1 and 2, and blue line, the sum of reactions 7 and 8. The rates of reaction 6, not shown in the plot, lie below $10^{-4}$.

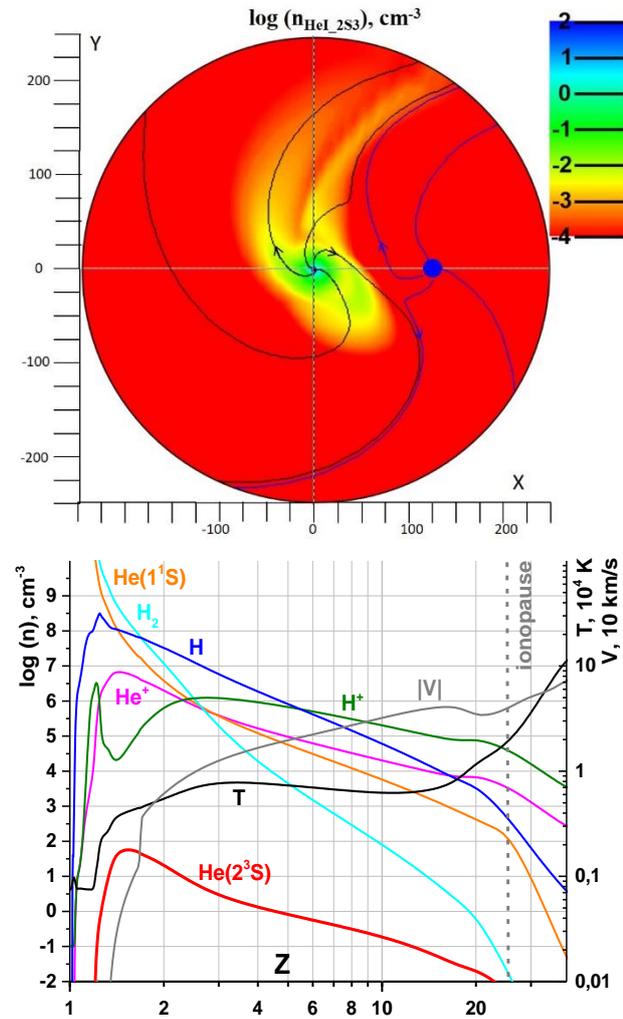

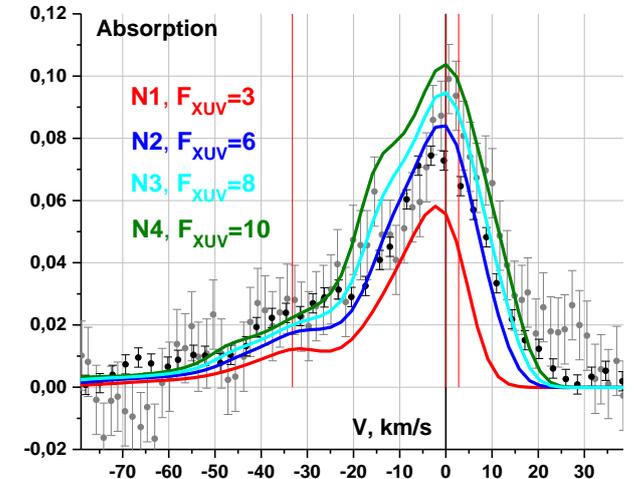

**Figure 3.** Total HeI($2^3$S) triplet absorption profiles, simulated with different $F_{XUV}$. Here and further on in similar plots the filled circles with error bars reproduce measurements from *Allart et al. 2019* (gray) and *Kirk et al. 2020* (black); red vertical lines indicate the positions of individual lines in the triplet.

**Figure 1.** Model run N1. *Top panel:* HeI($2^3$S) density distribution in orbital plane. The planet is at the center of coordinate (0,0) and moves anti-clockwise relative the star located at (120,0). Proton fluid streamlines are shown originated from planet (black) and from star (blue). *Bottom panel:* temperature and velocity profiles of HeI($2^3$S) (right axis) and density of major species (left axis) along the polar axis Z. The vertical dashed line indicates the position of boundary between the planetary and SW plasmas. The axes, here and further on, are scaled in planetary radii $R_p$.

Figure 3 compares the measured and simulated total absorption by HeI($2^3$S) atoms as the sum of contributions at the position of each line of the

According to Figure 3, the increase of $F_{XUV}$ increases the HeI($2^3$S) absorption. This is because the increased amount of $He^+$ causes the production of more HeI($2^3$S) atoms via recombination (reaction 9). The increased $F_{XUV}$ also leads to higher velocity and temperature of the escaping atmosphere, and as consequence, to a wider spectral range of the absorption. The calculated (see e.g., *Shaikhislamov et al. 2018*) mass loss rates M´$_p$ (Table 1) are comparable to the typical hot Jupiter's figures. Note that the deviation from the linear energy limited approximation is within 40%. The absorption profiles simulated in model runs N2, N3, N4, i.e. for $F_{XUV}$ ~ 6-10 erg cm$^{-2}$ s$^{-1}$ at 1 a.u., are in good agreement with the observations, in terms of both, the amplitude and the absorption range.

To highlight the role of radiation pressure acting on the HeI($2^3$S) atoms, besides of the typical value $F_{10830}$ = 11 erg s$^{-1}$ cm$^{-2}$ Å$^{-1}$ at 1 a.u. (model run N2), we did simulations with lower and higher fluxes, while keeping other model parameters the same as in N2 (Figure 4, top panel). The increased acceleration of HeI($2^3$S), due to the increased radiation pressure, displaces the absorption profiles towards blue area of the Doppler sifted velocities. The model run N5, with the weakest radiation pressure, reveals an additional absorption feature in the red wing [10; 20] km/s. It is produced in a wide halo of ~10$R_p$, extending beyond the Roche lobe, where the HeI($2^3$S) atoms move also towards the star driven by stellar gravity.

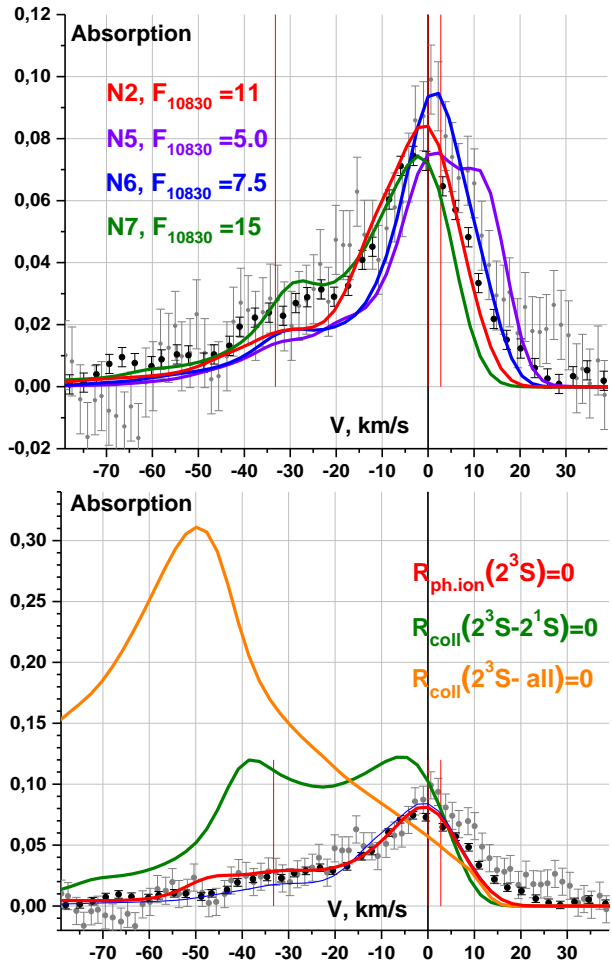

**Figure 4.** *Top panel:* Total HeI($2^3$S) triplet absorption profiles, simulated with different radiation pressures, taking the typical (N2) and the artificially prescribed values of $F_{10830}$. *Bottom panel:* Comparison of the absorption profile obtained in the model run N2 (blue line) with those obtained with the same parameters, as in N2, but without photoionization of HeI($2^3$S) (red line), without electron impact transfer $2^3$S – $2^1$S (green line), and without any collisional depopulation of HeI($2^3$S) (orange line).

The effect of radiation pressure acceleration is inhibited by the life time of metastable HeI($2^3$S) atoms until their destruction. To show this, we compare in bottom panel of Figure 4 the absorption profile, simulated in the model run N2, with those obtained for the same parameters, but after "switching-off" certain reactions, responsible for the destruction of HeI($2^3$S) atoms. Namely, the cases without photoionization of HeI($2^3$S) (reaction 2), or without electron impact transfer from the triplet to singlet sate (reaction 3), and without any collisional depopulation of HeI($2^3$S) metastable state (reactions 3-8), were separately considered. As it can be seen, the photoionization of HeI($2^3$S) has practically no effect. At the same time, the absence of collisional depopulation of HeI($2^3$S) leads to a significantly broader absorption profile with a much stronger peak, as compared to the observations. This effect is related with the strong impact of radiation pressure, acting on the HeI($2^3$S) atoms. This particular case is similar to one considered in *Allart et al. (2019)*, who also did not take into account the HeI($2^3$S) collisions, while considering a similar set of other parameters of the system. The increased absorption, uninhibited by collisions, led *Allart et al. (2019)* to assume a significantly reduced stellar NIR flux, in order to fit the simulations with measurements. It has to be noted that the radiation self-shielding around 10830 Å is not taken into account in our modelling. However, it takes place only very close to the planet, where the contribution of radiation pressure is not crucial.

Detailed analysis of simulations shown in Figure 3 reveals that the HeI($2^3$S) absorption around the line center ±10 km/s originates in the optically thick ($\tau_{10830}$ ~1) region close to the planet (<2$R_p$), whereas the absorption in the blue wing [-40; -10] km/s is provided by atoms from the optically thin ($\tau_{10830}$<<1) envelope at the heights of (1.5-3.5)$R_p$, where the radiation pressure becomes important. Therefore, the absorption in blue wing of the profile is more sensitive to the helium abundance than that at the center. Since the simulated absorption profiles in Figure 3 are generally below the measurements in the blue wing, one may expect that some increase of He/H might improve the fit. On the other hand, the effect of a decreased He/H can be to certain extend compensated by the increased $T$ and $V$ of the escaping material, i.e. by increasing $F_{XUV}$. To verify these expectations, we did simulations with different values of He/H and $F_{XUV}$ (see in Figure 5). In case of $F_{XUV}$=3 erg cm$^{-2}$ s$^{-1}$ at 1 a.u. and He/H=0.25 the spectral profile lacks the absorption in the red wing, while for $F_{XUV}$=6 erg cm$^{-2}$ s$^{-1}$ and He/H=0.15 it shows too high absorption around the line center. Finally, the increased $F_{XUV}$=10 erg cm$^{-2}$ s$^{-1}$ and slightly decreased He/H=0.075 provide better correspondence between the simulation and measurements in both, the blue and red wings of the absorption profile. The corresponding reduced $\chi^2=(N-2)^{-1}\sum(F_i-S_i)^2/\sigma^2$ in the range [-50; 30] km/s for the Allart et al. 2019 dataset are indicated in Figure 5. Altogether, we can conclude that the stellar XUV flux of (6-10) erg cm$^{-2}$ s$^{-1}$ at 1 a.u. and the solar-like helium abundance He/H = (0.075…0.15), used in the model runs N3,N9,N10, provide the best fit to observations.

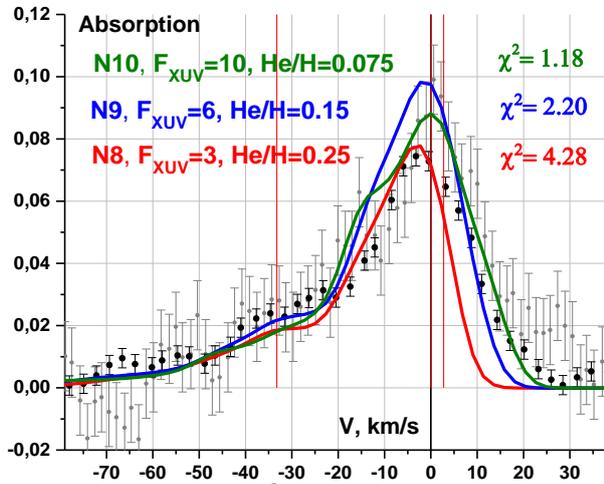

**Figure 5.** Total HeI($2^3$S) triplet absorption profiles, simulated with different He/H and $F_{XUV}$.

**4. Discussion and conclusion**

Based on the 3D HD multi-fluid modelling of the WASP-107 system, we reproduce the planetary atmospheric absorption at the position of metastable helium triplet at 10830 Å and constrain the values of the stellar XUV flux and planetary upper atmospheric helium abundance. WASP-107b appeared a first planet, for which the solar value of He/H is derived, in contrast to the previously analyzed planets, e.g., GJ3470b (*Ninan et al. 2020*, *Shaikhislamov et al. 2020c*) and HD209458b (*Lampon et al. 2020*), where the He/H values, obtained by similar model fitting to observations, were essentially sub-solar.

We confirm the finding of *Spake et al. (2018)* and *Allart et al. (2019)* that the radiation pressure acting on the metastable HeI($2^3$S) is an important factor for WASP-107b, affecting the shape of the absorption profiles. We show that the effect of the radiation pressure is counterbalanced by the processes of collisional depopulation of HeI($2^3$S) state, mainly by the $2^3$S – $2^1$S transition, while the impact of HeI($2^3$S) photoionization remains negligible.

Altogether, the HeI($2^3$S) absorption observed for WASP-107b can be well interpreted with the expected parameters of the system and appropriate account of the electron and atom impact processes, i.e., without invoking of any specific additional assumptions.

**Acknowledgements:**
This work was supported by RFBR grant № 18-12-00080. I.G.B, I.B.M. and M.A.E. also acknowledge RFBR grant 20-02-00520. M.L.K., I.F.S and M.S.R. are thankful to Russian ministry of science and higher education (contract 075-15-2020-780). M.L.K acknowledges projects I2939-N27, S11606-N16 of the Austrian Science Fund (FWF) and grant No.075-15-2019-1875 "Study of stars with exoplanets" from the government of Russian Federation. Parallel computing simulations for this study were done at Computation Center of Novosibirsk State University, Siberian Supercomputer Center of SB RAS and RAS Joint Supercomputer Center.

**Data availability:** The underlying data were published in the cited papers. Details of modelling code may be shared on reasonable request.